\documentclass{l4dc2027}

\makeatletter
\renewcommand{\ps@jmlrtps}{%
  \let\@mkboth\@gobbletwo
  \let\@oddhead\@empty
  \let\@evenhead\@empty
  \def\@oddfoot{\hfill\small\rmfamily\thepage\hfill}%
  \let\@evenfoot\@oddfoot
}
\makeatother

\title{Exploration as Feedback in Model Predictive Path Integral Control}

\usepackage{times}
\usepackage{mathtools}
\usepackage{wrapfig}
\mathtoolsset{showonlyrefs=false}

\newcommand{\R}{\mathbb{R}}
\newcommand{\E}{\mathbb{E}}
\newcommand{\ess}{\mathrm{ess}_\infty}
\newcommand{\Klq}{K_{\mathrm{lq}}}
\newcommand{\calA}{\mathcal{A}}
\newcommand{\calH}{\mathcal{H}}
\newcommand{\Sh}{\hat{S}}
\newcommand{\Mb}{\bar{M}}
\newcommand{\Nb}{\bar{N}}

\author{%
 \Name{Junsei Ito} \Email{distinction0625@moegi.waseda.jp}\\
 \addr Waseda University
 \AND
  \Name{Kohei Honda} \Email{honda.kohei.k3@f.mail.nagoya-u.ac.jp}\\
 \addr Nagoya University
 \AND
 \Name{Yasuaki Wasa} \Email{wasa@waseda.jp}\\
 \addr Waseda University%
}

\begin{document}

\maketitle

\begin{abstract}
This paper investigates how exploration affects closed-loop stability in model predictive path integral (MPPI) control. With a warm start, MPPI optimizes a control sequence by sampling candidates around the previous solution, and their covariance, the exploration covariance, is typically tuned to balance solution quality and computational cost. 
Small exploration is attractive because most samples then contribute to the solution, and existing stability analyses treat reduced exploration only as a bias toward the previous solution. Whether insufficient exploration can destabilize the closed loop has therefore remained open.
We show that, for linear quadratic (LQ) problems with infinitely many samples, the exploration covariance also determines the feedback gain, \textit{i.e.}, how strongly MPPI updates the previous solution in response to the current state. With small exploration at a fixed temperature, MPPI nearly repeats the previous solution and fails to stabilize an exponentially unstable plant. Because a single covariance governs both exploration and feedback, stability and sample efficiency are in structural conflict: stabilization requires large exploration, which leaves few samples that contribute to the solution.
To resolve this conflict, we propose Stabilizing Subspace MPPI (SS-MPPI), which computes the control inputs required for stabilization and samples only the rest. For LQ problems, SS-MPPI remains stable with small exploration and is thus sample-efficient. Simulations of an inverted pendulum and a quadruped robot confirm the analysis, showing that MPPI fails under small exploration, whereas SS-MPPI remains stable even with few samples.

\end{abstract}



\section{Introduction}
\label{sec:intro}

Is exploration in sampling-based model predictive control (MPC) only a means of finding better solutions? Exploration is usually regarded as such and is tuned in terms of solution quality and computational cost. Model predictive path integral (MPPI) control~\citep{WDG+16,WDG+18}, a representative sampling-based MPC method, solves a finite-horizon optimal control problem as probabilistic inference over control sequences. MPPI draws candidate control sequences from a Gaussian distribution centered at the nominal control sequence and updates the solution by weighting the candidates according to their costs. In this way, MPPI explores the space of control sequences to find a better solution than the nominal one. The covariance of this Gaussian distribution, which we call the \emph{exploration covariance}, determines how widely the candidates spread around the nominal control sequence and, at the same time, how far the solution moves away from the nominal one. Within the finite-horizon optimization at one time step, this covariance is merely a hyperparameter that tunes how widely MPPI explores and how large the update is~\citep{Honda26}.

MPPI, however, is usually implemented with a warm start~\citep{WDG+18}. MPPI performs one update per time step and shifts the updated sequence by one step to form the next nominal control sequence. The finite-horizon optimizations at successive time steps are therefore not independent, since each update starts from the previous solution. In effect, the warm start turns MPPI into a dynamical controller that carries the previous control sequence as its internal state and forms a closed loop with the plant. \emph{What, then, does exploration do in this closed loop across time steps?}

In this paper, we show that exploration in MPPI with a warm start is not only a search for solutions but also \emph{feedback} that stabilizes the closed loop. The exploration covariance determines which control sequences are explored at each time step. In a single optimization, the same covariance merely sets the size of the update; across time steps, however, this update corrects the nominal control sequence inherited from the previous time step, so its size acts as the feedback gain of the closed loop. A single covariance thus plays two roles at once; exploration within the horizon and feedback across time steps. Our central claim is that making this dual role explicit allows the stability and the sample efficiency of MPPI to be understood within a single framework. In particular, since one covariance carries both stabilization and exploration, the conflict between stability and sample efficiency is a structural problem, not a matter of hyperparameter tuning.

The first contribution is to show theoretically that the exploration covariance of MPPI determines the feedback gain of the dynamical controller formed by the warm start. We establish this for linear quadratic (LQ) problems in the limit of infinitely many samples, where the MPPI update can be expressed as an interpolation between the warm-started nominal sequence and the exact MPC solution, that is, the exact minimizer of the same finite-horizon problem. The interpolation gain is set by the exploration covariance, \textit{i.e.}, large exploration brings the update close to exact MPC, whereas small exploration leaves it close to the warm start. In particular, for an exponentially unstable plant, too little exploration leaves the feedback insufficient and destabilizes the closed loop. For a fixed covariance shape and temperature, stabilization therefore requires exploration above a threshold. This instability persists even with infinitely many samples and is thus distinct from sampling error.

The second contribution is to show that setting the exploration covariance involves a structural conflict between closed-loop stability and sample efficiency. Stabilization requires large exploration to strengthen the feedback. Enlarging the same covariance, however, causes the importance weights to degenerate. For LQ problems, we quantify this loss by an upper bound on the normalized effective sample size (ESS) expressed in terms of the feedback gain. Moreover, we show that the temperature enters both the closed loop and the bound only as a scale of the covariance, so tuning the temperature at a fixed feedback gain does not improve the bound.

The third contribution is to propose Stabilizing Subspace MPPI (SS-MPPI), which resolves the conflict between stability and sample efficiency. Rather than sampling the entire control sequence, SS-MPPI computes its components in the directions required for closed-loop stabilization and samples only the remaining directions, thereby separating stabilization from exploration. For LQ problems, an appropriate choice of this stabilizing subspace keeps the closed loop stable with small exploration and brings the normalized ESS close to one. We further extend this construction to nonlinear models and test SS-MPPI on an inverted pendulum on a cart and a quadruped robot. The results show that SS-MPPI keeps the closed loop stable with small exploration and few samples.

\begin{figure}[t]
\centering
\includegraphics[width=0.82\linewidth,bb=8.08 2.00 507.56 137.70]{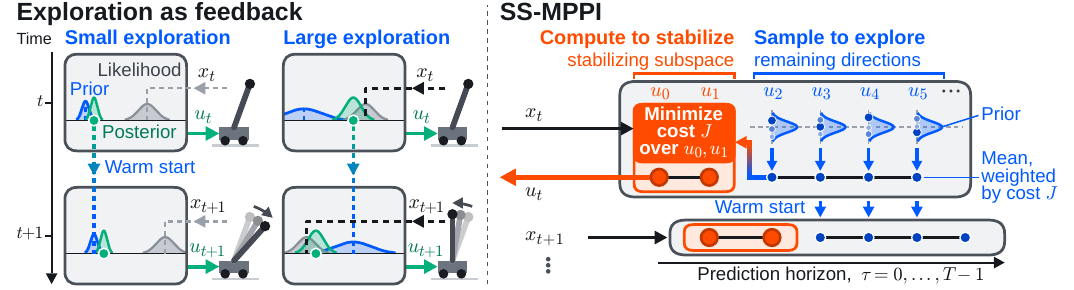}
\caption{Overview of exploration as feedback in MPPI (left) and the proposed SS-MPPI (right).}
\label{fig:teaser}
\end{figure}

\paragraph{Notation.}
For a vector $x$ and a matrix $X$, $\|x\|$ is the Euclidean norm, $\|X\|$ the spectral norm, $\rho(X)$ the spectral radius, and $X$ is Schur if $\rho(X)<1$.
$I$ (or $I_m$ for size $m$) and $0$ are the identity and zero matrices.
A map of class $C^k$ is $k$ times continuously differentiable.
$X\succeq0$ ($X\succ0$) means that $X$ is symmetric positive semidefinite (definite), $\mathcal{N}(\mu,\Sigma)$ is the Gaussian distribution with mean $\mu$, covariance $\Sigma\succ0$, and density $\mathcal{N}(\,\cdot\,;\mu,\Sigma)$, and $\E_p$ is the expectation under a density $p$.

\section{Related Work}
\label{sec:related}

The closed-loop stability of MPPI with the warm start has been studied by \citet{YK26a,YK26b}, who establish practical stability with finitely many samples by treating the update as a perturbation of a stabilizing reference controller.
Their guarantee requires a low temperature and enough samples, and a small covariance only leaves a bias toward the warm start, so whether too little exploration destabilizes the closed-loop system has remained open.
Real-time MPC solvers have also been analyzed as closed-loop systems with the warm-started control sequence as a state \citep{DFA+05,ABRR17}, but their update has no exploration.

By contrast, the update of MPPI itself has been analyzed for a given control problem.
The update has been interpreted as preconditioned gradient descent \citep{FSW26} and as expectation--maximization \citep{WSF26}, and the suboptimality of the update has been quantified \citep{HMDR25}.
The covariance has been shaped to speed up the update \citep{YPH+24}, and the number of samples required has been bounded \citep{YTK+22}.
These analyses treat the exploration covariance as a free parameter of one update, so the two roles have been studied apart.

A third line of work combines sampling with computation.
The computed part is an auxiliary controller that guides the samples \citep{GVG+21}, a feedback policy applied within the rollout to mitigate the divergence of the samples for an unstable plant \citep{PYZ+26}, a local model of the cost or of the rollouts \citep{SNTPG+26,BZTM26}, or a low-dimensional parameterization of the control sequence \citep{YD24}.
In all of the above, the part to compute is chosen by the designer, the cost, or the model, whereas the proposed SS-MPPI chooses the subspace to compute by the stability of the closed-loop system of the plant and the warm start.

\section{Preliminary}
\label{sec:prelim}
We consider the finite-horizon optimal control problem
\begin{equation}
    \min_{u_{0:T-1}}\ J(x_{0:T},u_{0:T-1})\quad\mathrm{s.t.}\quad x_{\tau+1}=f(x_{\tau},u_{\tau}),\quad \tau=0,\dots,T-1,\quad x_0=\bar x,
    \label{eq:ocp}
\end{equation}
where the horizon is a positive integer $T$, the state $x_{\tau}\in\R^n$ evolves from the initial state $\bar x\in\R^n$ under the control input $u_{\tau}\in\R^m$ by a map $f:\R^n\times\R^m\to\R^n$, and the cost function $J:\R^{n(T+1)}\times\R^{mT}\to[0,+\infty]$ evaluates the state and control input sequences.
The trajectory in~\eqref{eq:ocp} is determined by $\bar x$ and the control sequence $U:=[u_0^\top,\dots,u_{T-1}^\top]^\top\in\R^d$, $d:=mT$, whose first control input is $E_0U$, $E_0:=[I_m\ 0\ \cdots\ 0]$.
The cost is thus a function $J(x,U)$ of the initial state $\bar x=x$ and $U$.
At each time step $t$, MPC solves~\eqref{eq:ocp} from the current state $\bar x=x_t$, applies the first control input of the solution to the plant, and repeats from the new state~\citep{RMD17}.
We call the controller that applies the first control input of an exact minimizer of $J(x_t,\cdot)$ \emph{exact MPC}.

\paragraph{MPPI.}
MPPI~\citep{WDG+18} instead approximates the minimization by sampling, without derivatives.
At the time step $t$, MPPI draws $K$ independent control sequences $U^{(k)}$ from the Gaussian prior $p=\mathcal{N}(\mu_{t|t-1},\Sigma)$ around the nominal control sequence $\mu_{t|t-1}\in\R^d$ with the exploration covariance $\Sigma\succ0$.
Each sample is weighted by the likelihood $w(U):=e^{-J(x_t,U)/\lambda}$ with a temperature $\lambda>0$, and the weighted average $\hat\mu_{t|t}:=\sum_k\bar w_kU^{(k)}$ with the normalized weights $\bar w_k:=w(U^{(k)})/\sum_jw(U^{(j)})$ estimates the optimal control sequence.
The prior and the likelihood define the optimal control distribution $\pi(U\mid x_t)\propto w(U)\,p(U)$, which we assume to be normalizable, with the mean
\begin{equation}
    \mu_{t|t}=\E_p[w(U)]^{-1}\E_p[w(U)\,U].
    \label{eq:mean}
\end{equation}

\paragraph{Warm start and sample efficiency.}

MPPI applies $u_t=E_0\hat\mu_{t|t}$ and warm-starts the next update at $\mu_{t+1|t}=S\hat\mu_{t|t}$, where the nilpotent shift matrix $S\in\R^{d\times d}$ gives $SU=[u_1^\top,\dots,u_{T-1}^\top,0]^\top$.
The subscripts $t|t$ and $t+1|t$ denote the control sequence after the update at the time step $t$ and before the next update.
MPPI carries over only the mean and resets the covariance to the design parameter $\Sigma$ at every time step.
With the plant equal to the model $f$, the closed loop has the state $(x_t,\mu_{t|t-1})$ and closes through the warm start as well as through the plant, and as $K\to\infty$ it becomes the deterministic system $x_{t+1}=f(x_t,E_0\mu_{t|t})$, $\mu_{t+1|t}=S\mu_{t|t}$.
For finite $K$, the estimate degenerates when a few samples carry most of the weight~\citep{APSAS17}. We measure the sample efficiency by the almost sure limit of the normalized effective sample size (ESS) $1/(K\sum_k\bar{w}_k^2)$,
\begin{equation}
    \ess:=\E_p[w^2]^{-1}\E_p[w]^2.
    \label{eq:ess}
\end{equation}
Since the prior is both the proposal and a factor of the target $\pi$, $\Sigma$ changes the target, not only the proposal.
How the exploration covariance $\Sigma$ determines the stability of this closed loop, and how this relates to the sample efficiency~\eqref{eq:ess}, has not been characterized~\citep{Honda26}.

\section{Stability and Sample Efficiency of MPPI for LQ Problems}
\label{sec:analysis}

To characterize how the exploration covariance determines the closed-loop stability and the sample efficiency of MPPI, we first consider LQ problems.
For a linear plant with quadratic costs, the cost $J(x,U)$ is quadratic in $U$, so the mean~\eqref{eq:mean} has a closed form and the closed-loop system becomes explicit.
This LQ instance of~\eqref{eq:ocp} is also the model of a smooth problem near an equilibrium, which Section~\ref{sec:nonlinear} uses to treat nonlinear plants.
In this instance, the cost function and the map in~\eqref{eq:ocp} are
\begin{equation}
    J(x_{0:T},u_{0:T-1})=\sum_{\tau=0}^{T-1}(x_{\tau}^\top Qx_{\tau}+u_{\tau}^\top Ru_{\tau})+x_T^\top Px_T,\qquad f(x,u)=Ax+Bu,
    \label{eq:lq}
\end{equation}
where the weights $Q\in\R^{n\times n}$ and $R\in\R^{m\times m}$ satisfy $Q\succeq0$ and $R\succ0$, the matrices $A\in\R^{n\times n}$ and $B\in\R^{n\times m}$ form a stabilizable pair $(A,B)$, the pair $(A,Q)$ is detectable, and the terminal weight $P=P^\top\succeq0$ is the stabilizing solution of the discrete algebraic Riccati equation (DARE) $P=Q+A^\top PA-A^\top PB(R+B^\top PB)^{-1}B^\top PA$.
The terminal weight from the DARE is the standard choice of stabilizing MPC \citep{MRRS00}, under which the finite-horizon problem reproduces the infinite-horizon LQ regulator for every horizon.
The horizon therefore plays no role in the stability of exact MPC, and any instability of MPPI comes from its update.

In this setting, exact MPC is the LQ regulator.
The predicted state trajectory is linear in the state $x$ and the control sequence $U$, so the cost~\eqref{eq:lq} is quadratic in $U$,
\begin{equation}
    J(x,U)=U^\top MU+2U^\top Nx+c(x),
    \label{eq:quad}
\end{equation}
where $M\succ0$ and $N$ are given in Appendix~\ref{app:lifted} and $c(x)$ does not depend on $U$.
We call $M$ the curvature, the control-side counterpart of the posterior curvature in estimation, which decides how sharply the measurements determine the estimate \citep{IW26a}.
The minimizer of~\eqref{eq:quad} is $U^\star(x)=-Fx$ with the gain $F:=M^{-1}N$.
The first input of $U^\star(x)$ is that of the LQ regulator for every $T\ge1$, $E_0F=\Klq:=(R+B^\top PB)^{-1}B^\top PA$ (Appendix~\ref{app:riccati}).
Exact MPC is thus the regulator $u_t=-\Klq x_t$, and $A-B\Klq$ is Schur \citep{MRRS00}.

To characterize the closed-loop system of MPPI, we derive the mean~\eqref{eq:mean}.
Completing the square in~\eqref{eq:quad} gives $J(x,U)=(U-U^\star(x))^\top M(U-U^\star(x))+x^\top Px$ (Appendix~\ref{app:riccati}), so the likelihood $w$ is a Gaussian function centered at the control sequence $U^\star(x_t)$ of exact MPC with the precision $\alpha M$, where $\alpha:=2/\lambda$.
With the prior precision $\Sigma^{-1}$, the mean~\eqref{eq:mean} is therefore the precision-weighted average of the nominal control sequence $\mu_{t|t-1}$ and $U^\star(x_t)$, written as the update
\begin{equation}
    \mu_{t|t}=\mu_{t|t-1}+(I-\Gamma)\big(U^\star(x_t)-\mu_{t|t-1}\big),\qquad \Gamma:=(I+\alpha\Sigma M)^{-1},
    \label{eq:update}
\end{equation}
where the weight $\Gamma=(\Sigma^{-1}+\alpha M)^{-1}\Sigma^{-1}$ of $\mu_{t|t-1}$ is the share of the prior precision in the total precision.
We call $\mu_{t|t-1}$ the memory, since it carries the past through the warm start.
The update feeds back the gap between the memory and exact MPC with the gain $I-\Gamma$, which we call the feedback gain, so MPPI follows its memory for a small covariance and exact MPC for a large one.

The covariance counts as small or large only relative to the curvature, since $\Gamma$ depends on the product $\alpha\Sigma M$ alone.
To compare the two precisions by direction, we whiten the prior by the change of coordinates $\Sigma^{-1/2}U$, in which the prior has the precision $I$, the likelihood has the precision $\Sh:=\alpha\Sigma^{1/2}M\Sigma^{1/2}$, and $\Gamma$ becomes $(I+\Sh)^{-1}$.
In the eigendirection of $\Sh$ with the eigenvalue $s_i$, the likelihood is $s_i$ times as precise as the prior, so exact MPC and the memory have the weights $\kappa_i:=s_i/(1+s_i)$ and $1-\kappa_i$.
We call $\kappa_i$ the regulator weights since exact MPC is the regulator.

In this setting, MPPI with the warm start is a linear dynamical controller.
Since $U^\star(x_t)=-Fx_t$, the mean~\eqref{eq:update} is $\mu_{t|t}=\Gamma\mu_{t|t-1}-\Lambda x_t$ with the gain $\Lambda:=(I-\Gamma)F$, so the exploration covariance sets the gains of the controller through the weight $\Gamma$.
Combining the mean with the plant $x_{t+1}=Ax_t+BE_0\mu_{t|t}$ and the warm start $\mu_{t+1|t}=S\mu_{t|t}$, we obtain the closed-loop system of the plant and MPPI with infinitely many samples, the system of Section~\ref{sec:prelim} in this setting,
\begin{equation}
    \xi_{t+1}=\calA(\Sigma)\xi_t,\qquad
    \calA(\Sigma):=\begin{bmatrix}A-BE_0\Lambda & BE_0\Gamma\\ -S\Lambda & S\Gamma\end{bmatrix},
    \label{eq:Acl}
\end{equation}
where the augmented state $\xi_t:=(x_t,\mu_{t|t-1})\in\R^{n+d}$ collects the plant state and the memory.

\begin{definition}
\label{def:stability}
The closed-loop system~\eqref{eq:Acl} is exponentially stable if there are $c\ge1$ and $\beta\in(0,1)$ such that $\|\xi_t\|\le c\beta^t\|\xi_0\|$ for all $\xi_0$ and $t\ge0$ \citep[Appendix~B]{RMD17}.
\end{definition}

For the linear system~\eqref{eq:Acl}, exponential stability is equivalent to $\rho(\calA(\Sigma))<1$, and $\rho(\calA(\Sigma))>1$ implies instability.
The following theorem shows how the exploration covariance sets the feedback gain of the closed-loop system~\eqref{eq:Acl}.

\begin{theorem}
\label{thm:closedloop}
Consider the closed-loop system~\eqref{eq:Acl} with $\Sigma=\eta\Sigma_0$, where the shape $\Sigma_0\succ0$ is fixed and the scale $\eta>0$ is the exploration.
Then the following statements hold.
\begin{enumerate}
    \item[(a)] The limits of $\calA(\eta\Sigma_0)$ are
    \begin{equation}
        \lim_{\eta\to0}\calA(\eta\Sigma_0)=\begin{bmatrix}A & BE_0\\ 0 & S\end{bmatrix},\qquad
        \lim_{\eta\to\infty}\calA(\eta\Sigma_0)=\begin{bmatrix}A-B\Klq & 0\\ -SF & 0\end{bmatrix},
        \label{eq:limits}
    \end{equation}
    whose spectra are those of $A$ and of $A-B\Klq$, each together with zeros.
    \item[(b)] If $\rho(A)>1$, there exist $0<\eta_1\le\eta_2$ with the closed-loop system~\eqref{eq:Acl} unstable for $\eta<\eta_1$ and stable for $\eta>\eta_2$.
\end{enumerate}
\end{theorem}

\begin{proof}[Sketch]
(a) Every $s_i$ is proportional to $\eta$, so $\Gamma\to I$ and $\Lambda\to0$ as $\eta\to0$, and $\Gamma\to0$ and $\Lambda\to F$ as $\eta\to\infty$, where $E_0F=\Klq$.
Both limits are block triangular, and their lower-right blocks $S$ and $0$ are nilpotent.
(b) The spectral radius $\rho(\calA(\eta\Sigma_0))$ is continuous in $\eta$ and tends to $\rho(A)>1$ as $\eta\to0$ and to $\rho(A-B\Klq)<1$ as $\eta\to\infty$ (Appendix~\ref{app:thm1}).
\end{proof}

Theorem~\ref{thm:closedloop} shows that the exploration covariance moves the controller from replaying its memory with the plant in open loop to the regulator $u_t=-\Klq x_t$.
For $\rho(A)>1$, the exploration must therefore exceed a threshold, although the stable set need not be an interval since the spectral radius need not be monotone in $\eta$.
At each state and memory, more samples only bring the update of MPPI closer to the closed-loop system~\eqref{eq:Acl}, which is unstable below the threshold (Appendix~\ref{app:thm1}).

The same covariance also sets the sample efficiency, as the following proposition shows.

\begin{proposition}
\label{prop:inference}
At the augmented state $\xi_t$, the ESS~\eqref{eq:ess} satisfies
\begin{equation}
    \ess\le\textstyle\prod_{i=1}^d(1-\kappa_i^2)^{1/2},
    \label{eq:essclosed}
\end{equation}
with equality if and only if the memory $\mu_{t|t-1}$ equals the control sequence $U^\star(x_t)$ of exact MPC.
\end{proposition}

\begin{proof}[Sketch]
Both moments in~\eqref{eq:ess} are Gaussian integrals (Appendix~\ref{app:ess}).
Their determinant factors give $\prod_i(1+2s_i)^{1/2}/(1+s_i)$, which is the product in~\eqref{eq:essclosed} since $1+2s=(1+s)^2-s^2$, and their exponents give a factor at most one, with equality if and only if $\mu_{t|t-1}=U^\star(x_t)$.
\end{proof}

Theorem~\ref{thm:closedloop} and Proposition~\ref{prop:inference} show that the feedback and the inference pull the exploration covariance in opposite directions, which we call the conflict between stability and sample efficiency.
Closed forms of the ESS are known for Gaussian importance sampling \citep{APSAS17}, and the point here is that the factors of the bound~\eqref{eq:essclosed} are set by the regulator weights of Theorem~\ref{thm:closedloop}.
Enlarging the covariance raises every weight $\kappa_i$, which a plant with $\rho(A)>1$ needs by Theorem~\ref{thm:closedloop}(b), and lowers every factor $(1-\kappa_i^2)^{1/2}$, which vanishes as $\kappa_i\to1$.
The loss of ESS at the threshold depends on the plant and can be nearly total (Appendix~\ref{app:linear}).
For an isotropic covariance, the weights are largest in the stiff directions of the cost, the eigendirections of $M$ with large eigenvalues, so the ESS collapses there first.
Neither more samples nor the temperature lifts the bound~\eqref{eq:essclosed} at a fixed gain, since the temperature enters~\eqref{eq:Acl} and~\eqref{eq:essclosed} only through the product $\alpha\Sigma$ (Appendix~\ref{app:ess}).

\section{Stabilizing Subspace MPPI}
\label{sec:collapsed}

To resolve the conflict shown in Section~\ref{sec:analysis}, this section proposes Stabilizing Subspace MPPI (SS-MPPI), as shown in Figure~\ref{fig:teaser}.
In the directions where stability requires large regulator weights, SS-MPPI minimizes the cost instead of sampling, so that these weights no longer reduce the  ESS.
Formally, SS-MPPI places an improper flat prior proportional to one on a subspace containing these directions, the stabilizing subspace, integrates over the subspace analytically, and samples the rest with a small covariance.
By Theorem~\ref{thm:closedloop}, the flat prior, the limit of large exploration, makes the stabilizing subspace carry the feedback, and the computed coordinates drop out of the bound~\eqref{eq:essclosed}.

An orthogonal matrix $V=[V_h\ V_s]$ splits the control sequence as $U=V_hz_h+V_sz_s$ into the computed coordinates $z_h\in\R^{d_h}$ in the stabilizing subspace $\calH:=\operatorname{im}V_h$ and the sampled coordinates $z_s\in\R^{d_s}$, with $d_s:=d-d_h$.
In Section~\ref{sec:collapsedlq}, we specialize $\calH$ to the span of the first few control inputs, as illustrated in Figure~\ref{fig:teaser}.
SS-MPPI computes the mean of
\begin{equation}
    \pi_\calH(z_h,z_s\mid x_t)\propto e^{-J(x_t,V_hz_h+V_sz_s)/\lambda}\,\mathcal{N}(z_s;m_{t|t-1},\Sigma_s),
    \label{eq:collapsed}
\end{equation}
in which only the sampled coordinates have a Gaussian prior, with the exploration covariance \mbox{$\Sigma_s\succ0$} around their nominal value $m_{t|t-1}$ replacing $\mu_{t|t-1}$ since the flat prior has no center.
The controller applies $u_t=E_0U_{t|t}$ from the mean $U_{t|t}$ and keeps $m_{t+1|t}=V_s^\top SU_{t|t}$ as the memory.

\subsection{Linear quadratic problems}
\label{sec:collapsedlq}

In the LQ setting of Section~\ref{sec:analysis}, the integral over the computed coordinates has a closed form (Appendix~\ref{app:collapsedmean}).
The integral leaves $e^{-\bar J(x,z_s)/\lambda}$ up to a constant factor, so each sample is weighted by the profiled cost $\bar J(x,z_s):=\min_{z_h}J(x,V_hz_h+V_sz_s)$, whose curvature $\Mb\succ0$ is the Schur complement of the block $M_h:=V_h^\top MV_h$ of $V^\top MV$.
The control sequence $V_hz_h^\star(x,z_s)+V_sz_s$ at the minimizer $z_h^\star(x,z_s)$ in $\bar J$ is affine in $(x,z_s)$, and we write it as $-L_hx+T_sz_s$ with the gain $L_h:=V_hM_h^{-1}V_h^\top N$ and the lifting $T_s:=V_s-V_hM_h^{-1}V_h^\top MV_s$.
The means of $z_s$ and $U$ are then
\begin{equation}
    m_{t|t}=m_{t|t-1}+(I-\Gamma_s)\big(V_s^\top U^\star(x_t)-m_{t|t-1}\big),\qquad U_{t|t}=-L_hx_t+T_s m_{t|t},
    \label{eq:collapsedupdate}
\end{equation}
with $\Gamma_s:=(I+\alpha\Sigma_s\Mb)^{-1}$, so $m_{t|t}$ is the mean~\eqref{eq:update} for the sampled coordinates.
The augmented state is $\zeta_t:=(x_t,m_{t|t-1})$.
The following theorem shows that a condition on the stabilizing subspace alone gives stability at every sufficiently small exploration, where the ESS tends to one.

\begin{theorem}
\label{thm:collapsed}
Consider the closed-loop system of the linear plant of Section~\ref{sec:analysis} and SS-MPPI with the mean~\eqref{eq:collapsedupdate}.
Then the following statements hold (proof in Appendix~\ref{app:thm3}).
\begin{enumerate}
    \item[(a)] The closed-loop system is $\zeta_{t+1}=\calA_s(\Sigma_s)\zeta_t$, where $\calA_s(\Sigma_s)$ is the matrix~\eqref{eq:A0} below with $L_h$ replaced by $L_h+T_s(I-\Gamma_s)V_s^\top F$ and $T_s$ by $T_s\Gamma_s$, which tends as $\Sigma_s\to0$ to
    \begin{equation}
        \calA_0:=\begin{bmatrix}A-BE_0L_h & BE_0T_s\\ -V_s^\top SL_h & V_s^\top ST_s\end{bmatrix}.
        \label{eq:A0}
    \end{equation}
    If $\calA_0$ is Schur, there is $\varsigma>0$ such that $\calA_s(\Sigma_s)$ is Schur for every $\Sigma_s\succ0$ with $\|\Sigma_s\|<\varsigma$.
    \item[(b)] At the augmented state $\zeta_t$, the asymptotic ESS~\eqref{eq:ess} of the sampled coordinates satisfies~\eqref{eq:essclosed} with $d$ and $\Sh$ replaced by $d_s$ and the whitened curvature $\Sh_s:=\alpha\Sigma_s^{1/2}\Mb\Sigma_s^{1/2}$.
    Equality holds if and only if $m_{t|t-1}=V_s^\top U^\star(x_t)$.
    The bound is thus~\eqref{eq:essclosed} with the regulator weights of the sampled coordinates, and the ESS tends to one as $\Sigma_s\to0$ at a fixed $\zeta_t$.
\end{enumerate}
\end{theorem}

By Theorem~\ref{thm:collapsed}, the conflict disappears when $\calA_0$ is Schur, since stability then requires no exploration.
A natural candidate for the stabilizing subspace is the span of the top eigenvectors of $M$, but this span need not make $\calA_0$ Schur even for $T=1$.
The following proposition gives the design principle, since a shift-invariant subspace makes the memory harmless (proof in Appendix~\ref{app:shift}).

\begin{proposition}
\label{prop:shift}
If $S\calH\subseteq\calH$, then $\calA_0$ is block upper triangular with the nilpotent memory block $V_s^\top SV_s$, and $\rho(\calA_0)=\rho(A-BE_0L_h)$.
\end{proposition}

We take the span of the first $k_b$ input blocks, which is shift-invariant for every $k_b$, with the smallest $k_b\in\{1,\dots,T\}$ for which $\rho(A-BE_0L_h)<1$.
Such $k_b$ exists, since $k_b=T$ gives $L_h=F$ and $A-BE_0F=A-B\Klq$, and SS-MPPI is then exact MPC.

\subsection{Extension to nonlinear plants}
\label{sec:nonlinear}

\begin{algorithm2e}[t]
\caption{Stabilizing Subspace MPPI (SS-MPPI)}
\label{alg:collapsed}
\KwIn{linearization $(A,B)$, weights $Q$, $R$, horizon $T$, variance $\sigma_s^2$, temperature $\lambda$, samples $K$}
\textbf{Offline:} compute $P$ by the DARE, $M$, and $N$, and let $\calH$ be the span of the first $k_b$ input blocks with the smallest $k_b$ such that $\rho(A-BE_0L_h)<1$\;
\textbf{Online, at each time step $t$:} compute $r_0$ and $H$ of~\eqref{eq:gn} at $(x_t,\mu_{t|t-1})$, and for $k=1,\dots,K$, draw $\delta_s^{(k)}\sim\mathcal N(0,\sigma_s^2I_{d_s})$, compute $\delta_h^{(k)}$ by~\eqref{eq:gn}, and set $U^{(k)}:=\mu_{t|t-1}+V_s\delta_s^{(k)}+V_h\delta_h^{(k)}$\;
set $\hat\mu_{t|t}:=\sum_k\bar w_kU^{(k)}$ with $\bar w_k$ as in Section~\ref{sec:prelim}, apply $u_t:=E_0\hat\mu_{t|t}$, and set $\mu_{t+1|t}:=S\hat\mu_{t|t}$\;
\end{algorithm2e}

In this subsection, we extend SS-MPPI to the nonlinear model $f$ of Section~\ref{sec:prelim} and show that the stability condition of Section~\ref{sec:collapsedlq} transfers to the closed-loop system near an equilibrium.
The integral over the computed coordinates then has no closed form, so SS-MPPI replaces the marginal of~\eqref{eq:collapsed} by the profile, weighting each sample by the cost at the minimizer $z_h^\star$, and approximates the minimizer by one Gauss--Newton step.
For this step, the cost is $J(x,U)=\|\mathcal{R}(x,U)\|^2$, where the residual map $\mathcal{R}:\R^n\times\R^d\to\R^q$ is of class $C^2$ and stacks the residuals of the states and the inputs along the predicted state trajectory.
The residual is linearized at the full warm start $\mu_{t|t-1}$, which SS-MPPI therefore keeps as the memory.
With $r_0:=\mathcal{R}(x_t,\mu_{t|t-1})$ and $H:=\partial\mathcal{R}/\partial U(x_t,\mu_{t|t-1})$, each sample consists of a perturbation $\delta_s\sim\mathcal{N}(0,\Sigma_s)$ and the Gauss--Newton step
\begin{equation}
    \delta_h=-(H_h^\top H_h)^{-1}H_h^\top\big(r_0+H_s\delta_s\big),\qquad H_h:=HV_h,\qquad H_s:=HV_s,
    \label{eq:gn}
\end{equation}
and the sample $U=\mu_{t|t-1}+V_s\delta_s+V_h\delta_h$ is weighted by its cost $J(x_t,U)$.
In the LQ setting of Section~\ref{sec:analysis}, where $\mathcal{R}$ is affine in $U$, the step~\eqref{eq:gn} is exact, and the controller with infinitely many samples is the mean of~\eqref{eq:collapsed} (Appendix~\ref{app:local}).
The controller takes the stabilizing subspace and the terminal weight from a linearization (Algorithm~\ref{alg:collapsed}).
The following proposition transfers the stability condition of Theorem~\ref{thm:collapsed} at the linearization to the nonlinear closed-loop system.

\begin{proposition}
\label{prop:local}
Let the map $f$ in~\eqref{eq:ocp} be of class $C^1$ with $f(0,0)=0$ and $\mathcal{R}(0,0)=0$, let $H_h$ have full column rank at the origin, and let $J$ and $\nabla_{(x,U)}J$ grow at most polynomially in $U$, uniformly for $x$ near the origin.
Form $M=H^\top H$, $N=H^\top L$, $L_h$, and $T_s$ from the Jacobians $H$ and $L$ of $\mathcal{R}$ in $U$ and $x$ at the origin, and $\calA_0$ by~\eqref{eq:A0} with the Jacobians $A$ and $B$ of $f$ at the origin.
If $\rho(\calA_0)<1$, there is $\varsigma>0$ such that, for every $\Sigma_s\succ0$ with $\|\Sigma_s\|<\varsigma$, the closed-loop system of the plant $f$ and Algorithm~\ref{alg:collapsed} with infinitely many samples has an exponentially attracting fixed point of $(x_t,\mu_{t|t-1})$ that tends to the origin as $\Sigma_s\to0$.
\end{proposition}

The proof is in Appendix~\ref{app:local}.
At zero exploration, SS-MPPI is a deterministic Gauss--Newton controller whose Jacobian at the origin has the diagonal blocks $\calA_0$ and $0$, the latter because the step overwrites the computed coordinates of the memory.
The weighted update is a $C^1$ perturbation of the Gauss--Newton controller as $\Sigma_s\to0$.
The fixed point may differ from the origin because, for a nonquadratic cost, the weighted mean of the samples need not be the minimizer.

\begin{figure}[t]
\centering
\includegraphics[width=0.79\linewidth,bb=0 0 515.52 144]{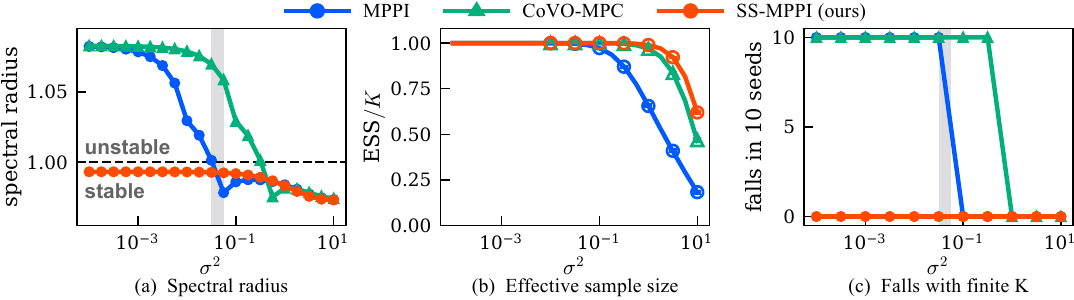}
\caption{Inverted pendulum on a cart versus the exploration variance.
(a) Linearization, threshold of MPPI in the grey band.
(b) Linearization at zero gap, $\mu_{t|t-1}=U^\star(x_t)$, predicted (lines) and Monte Carlo estimates (markers).
(c) Nonlinear model with $K=256$.}
\label{fig:cartpole}
\end{figure}

\section{Numerical Experiments}
\label{sec:numerical}
This section verifies the analysis and evaluates SS-MPPI with finitely many samples.
On an inverted pendulum on a cart, near the upright equilibrium where the theory applies, we verify the threshold of Theorem~\ref{thm:closedloop}, the bound of Proposition~\ref{prop:inference}, and the stability of SS-MPPI at small exploration.
On a Go2 quadruped, we test whether the subspace selected at standing stabilizes walking with few samples.
We compare SS-MPPI with two baselines: MPPI with $\Sigma=\sigma^2I$ and CoVO-MPC \citep{YPH+24}, which shapes the covariance as $\Sigma\propto M^{-1/2}$ with the determinant of $\sigma^2I$, to test whether shaping the covariance alone resolves the conflict.
All three use the same cost, temperature, horizon, and DARE terminal weight at the linearization, with the models and settings in Appendices~\ref{app:linear} and~\ref{app:go2}.

\paragraph{Inverted pendulum.}
The control objective is to stabilize the pendulum at its upright equilibrium, whose linearization has the unstable pole $1.08$.
To exclude sampling error and nonlinearity, we first sweep $\sigma^2$ from $10^{-4}$ to $10$ on this linearization and compute the spectral radius of~\eqref{eq:Acl} and the bound~\eqref{eq:essclosed} at zero gap, which matches Monte Carlo to one percent (Figure~\ref{fig:cartpole}(a,b)). MPPI is stable only above the threshold $\sigma^2=0.033$, as Theorem~\ref{thm:closedloop}(b) predicts, while its ESS falls below one half at $\sigma^2=2.0$. CoVO-MPC raises the ESS but moves the threshold tenfold to $\sigma^2=0.33$, so shaping the covariance trades stability for sample efficiency. SS-MPPI, which computes the first input ($k_b=1$), is stable throughout with the ESS above $0.6$, so a single computed input removes the threshold.

To test whether the threshold persists with finitely many samples, we next run the nonlinear model for 1000 time steps from an initial angle of \(0.05\) rad over ten seeds and count the failures, defined as \(|\theta|>\pi/2\) (Figure~\ref{fig:cartpole}(c)). MPPI fails for all ten seeds when \(\sigma^2\le 0.032\) and for none when \(\sigma^2\ge 0.1\), for \(K=64\), \(256\), and \(1024\). The threshold thus persists, and more samples do not compensate for insufficient exploration. CoVO-MPC with \(K=256\) fails for \(\sigma^2\) up to \(0.32\), again close to its predicted threshold, whereas SS-MPPI with \(K=256\) does not fail at any tested value.

\begin{figure}[t]
\centering
\includegraphics[width=0.77\linewidth,bb=0 0 515.52 132]{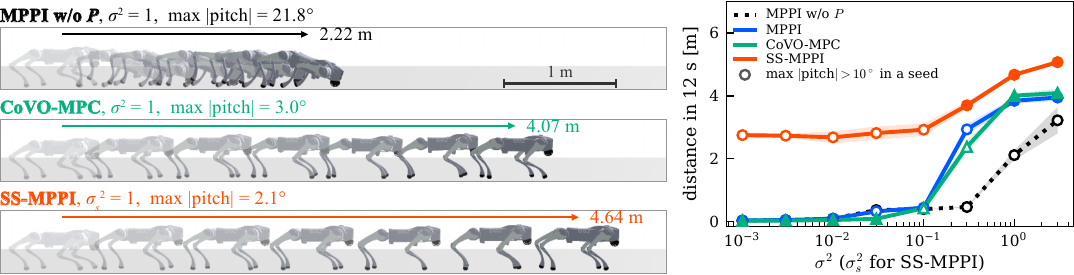}
\caption{Go2 commanded to walk at $0.5$ m/s.
Left: seed $0$ every $2$ s at $\sigma^2=\sigma_s^2=1$ and $K=1024$.
Right: distance in $12$ s vs.\ the variance (mean $\pm$ s.d.), hollow if the pitch exceeded $10^\circ$.}
\label{fig:go2}
\end{figure}

\paragraph{Quadruped.}
To test whether SS-MPPI maintains closed-loop stability far from equilibrium, where Proposition~\ref{prop:local} provides no guarantee, we make a Unitree Go2 walk forward at $0.5$ m/s in the Genesis simulator \citep{Genesis24}, using the prediction model and cost described in Appendix~\ref{app:go2}.
Because walking has no fixed equilibrium, the stabilizing subspace and terminal weight are computed once from the linearization about the standing configuration, for which Algorithm~\ref{alg:collapsed} selects the first two input blocks.
We run each controller for $12$ s over five seeds and measure the distance traveled and the maximum pitch, which we use as an indicator of instability, while sweeping the exploration variance and the number of samples (Figure~\ref{fig:go2} and Table~\ref{tab:go2}). 

MPPI without the terminal weight (denoted MPPI w/o $P$) pitches forward.
The terminal weight enables MPPI to walk steadily, with the pitch below $10^\circ$, at $\sigma^2=1$, but reduces the ESS at $\sigma^2=0.1$ from $0.63$ to $0.012$. Since $P$ increases the curvature $M$, the conflict of Section~\ref{sec:analysis} reappears in walking. MPPI and CoVO-MPC stand still or fall for $\sigma^2\le0.1$ and walk steadily only for $\sigma^2\ge1$.
SS-MPPI walks for every variance from $10^{-3}$ to $3$, walks steadily from $0.3$ to $3$, and covers up to $5.1$ m, so the subspace selected at standing removes the threshold in walking within the tested range.

\begin{wraptable}{r}{0.555\linewidth}
\vspace{-1.2em}
\centering
\caption{Go2 at $\sigma^2=\sigma_s^2=1$ over five seeds (mean $\pm$ s.d.). SS-MPPI with $k_b=T$ or $\Sigma_s=0$ does not sample, so the five runs coincide.}
\label{tab:go2}
\footnotesize
\setlength{\tabcolsep}{2pt}
\begin{tabular}{lcccc}
\hline
& \multicolumn{2}{c}{$K=1024$} & \multicolumn{2}{c}{$K=16$} \\
& distance $\uparrow$ & pitch $\downarrow$ & distance $\uparrow$ & pitch $\downarrow$ \\
& [m] & [$^\circ$] & [m] & [$^\circ$] \\
\hline
MPPI w/o $P$ & $2.12{\pm}0.16$ & $22.3{\pm}0.4$ & $1.57{\pm}0.11$ & $27.3{\pm}2.5$ \\
MPPI & $3.85{\pm}0.10$ & $4.7{\pm}0.8$ & $2.56{\pm}0.06$ & $10.4{\pm}1.7$ \\
CoVO-MPC & $4.01{\pm}0.05$ & $3.5{\pm}0.4$ & $2.56{\pm}0.13$ & $9.3{\pm}1.5$ \\
SS-MPPI, $k_b{=}T$ & $3.45$ & $4.0$ & $3.45$ & $4.0$ \\
SS-MPPI, $\Sigma_s{=}0$ & $2.54$ & $16.9$ & $2.54$ & $16.9$ \\
\hline
\textbf{SS-MPPI (ours)} & $\mathbf{4.68{\pm}0.04}$ & $\mathbf{2.2{\pm}0.4}$ & $\mathbf{4.77{\pm}0.13}$ & $\mathbf{5.0{\pm}1.2}$ \\
\hline
\end{tabular}
\vspace{-1em}
\end{wraptable}
Next, to test whether SS-MPPI needs fewer samples, we reduce $K$ at $\sigma^2=\sigma_s^2=1$, where all controllers with $P$ walk steadily. At $K=16$, MPPI and CoVO-MPC lose over a third of their distance and reach pitches of about $10^\circ$, whereas SS-MPPI maintains its distance at about $5^\circ$. This gain is not due to more rollouts: SS-MPPI with $K=16$ evaluates $257$ rollouts per step, yet walks farther than MPPI with $K=256$ ($3.46$ m) or $1024$, though at higher computational cost (Appendix~\ref{app:go2}). Without sampling, SS-MPPI walks only $2.54$ m, while computing the full sequence ($k_b=T$) yields $3.45$ m versus $4.68$ m with $K=1024$, so sampling the remaining coordinates is essential.

\section{Conclusion}
\label{sec:conclusion}
In this work, we have shown that exploration in MPPI with the warm start acts as feedback. For LQ problems with infinitely many samples, the exploration covariance sets the closed-loop feedback gain, so an exponentially unstable plant requires exploration above a threshold, whereas the same covariance lowers the bound on the ESS.
This conflict between stability and sample efficiency is structural, and more samples do not resolve it.
The proposed SS-MPPI resolves it by computing the control sequence in a stabilizing subspace and sampling the rest, and it kept the closed loop stable at small exploration and with few samples on an inverted pendulum and a quadruped robot.

\subsection*{Generative AI Use}
Anthropic's Claude was used to assist in writing the experiment code and to improve readability.
The authors reviewed and verified all AI-assisted content and take responsibility for the paper.

\acks{This work was partially supported by the Japan Science and Technology Agency (JST), ACT-X, Grant No.~JPMJAX25C3 and JST Moonshot R\&D Program Japan Grant No.~JPMJMS263D.}

\bibliography{refs}

\begin{thebibliography}{24}
\providecommand{\natexlab}[1]{#1}
\providecommand{\url}[1]{\texttt{#1}}
\expandafter\ifx\csname urlstyle\endcsname\relax
  \providecommand{\doi}[1]{doi: #1}\else
  \providecommand{\doi}{doi: \begingroup \urlstyle{rm}\Url}\fi

\bibitem[Agapiou et~al.(2017)Agapiou, Papaspiliopoulos, Sanz-Alonso, and
  Stuart]{APSAS17}
S.~Agapiou, O.~Papaspiliopoulos, D.~Sanz-Alonso, and A.~M. Stuart.
\newblock Importance sampling: {I}ntrinsic dimension and computational cost.
\newblock \emph{Statistical Science}, 32\penalty0 (3):\penalty0 405--431, 2017.

\bibitem[Allan et~al.(2017)Allan, Bates, Risbeck, and Rawlings]{ABRR17}
D.~A. Allan, C.~N. Bates, M.~J. Risbeck, and J.~B. Rawlings.
\newblock On the inherent robustness of optimal and suboptimal nonlinear {MPC}.
\newblock \emph{Systems \& Control Letters}, 106:\penalty0 68--78, 2017.

\bibitem[Barto et~al.(1983)Barto, Sutton, and Anderson]{BSA83}
A.~G. Barto, R.~S. Sutton, and C.~W. Anderson.
\newblock Neuronlike adaptive elements that can solve difficult learning
  control problems.
\newblock \emph{IEEE Transactions on Systems, Man, and Cybernetics},
  SMC-13\penalty0 (5):\penalty0 834--846, 1983.

\bibitem[Belvedere et~al.(2026)Belvedere, Ziegltrum, Turrisi, and
  Modugno]{BZTM26}
T.~Belvedere, M.~Ziegltrum, G.~Turrisi, and V.~Modugno.
\newblock {Feedback-MPPI}: {F}ast sampling-based {MPC} via rollout
  differentiation -- {A}dios low-level controllers.
\newblock \emph{IEEE Robotics and Automation Letters}, 11\penalty0
  (1):\penalty0 1--8, 2026.

\bibitem[Diehl et~al.(2005)Diehl, Findeisen, Allg{\"o}wer, Bock, and
  Schl{\"o}der]{DFA+05}
M.~Diehl, R.~Findeisen, F.~Allg{\"o}wer, H.~G. Bock, and J.~P. Schl{\"o}der.
\newblock Nominal stability of real-time iteration scheme for nonlinear model
  predictive control.
\newblock \emph{IEE Proceedings -- Control Theory and Applications},
  152\penalty0 (3):\penalty0 296--308, 2005.

\bibitem[Fazlyab et~al.(2026)Fazlyab, Sharifi, and Wang]{FSW26}
M.~Fazlyab, S.~Sharifi, and J.~Wang.
\newblock Model predictive path integral control as preconditioned gradient
  descent.
\newblock \emph{IEEE Control Systems Letters}, 10:\penalty0 2089--2094, 2026.

\bibitem[Florian(2007)]{Florian07}
R.~V. Florian.
\newblock Correct equations for the dynamics of the cart-pole system.
\newblock Technical report, Center for Cognitive and Neural Studies (Coneural),
  Cluj-Napoca, Romania, 2007.
\newblock First version 2005, updated February 2007.

\bibitem[Gandhi et~al.(2021)Gandhi, Vlahov, Gibson, Williams, and
  Theodorou]{GVG+21}
M.~S. Gandhi, B.~Vlahov, J.~Gibson, G.~Williams, and E.~A. Theodorou.
\newblock Robust model predictive path integral control: {A}nalysis and
  performance guarantees.
\newblock \emph{IEEE Robotics and Automation Letters}, 6\penalty0 (2):\penalty0
  1423--1430, 2021.

\bibitem[{Genesis Authors}(2024)]{Genesis24}
{Genesis Authors}.
\newblock Genesis: A generative and universal physics engine for robotics and
  beyond.
\newblock \url{https://github.com/Genesis-Embodied-AI/genesis-world}, December
  2024.

\bibitem[Homburger et~al.(2025)Homburger, Messerer, Diehl, and Reuter]{HMDR25}
H.~Homburger, F.~Messerer, M.~Diehl, and J.~Reuter.
\newblock Optimality and suboptimality of {MPPI} control in stochastic and
  deterministic settings.
\newblock \emph{IEEE Control Systems Letters}, 9:\penalty0 763--768, 2025.

\bibitem[Honda(2026)]{Honda26}
K.~Honda.
\newblock Model predictive control via probabilistic inference: {A} tutorial
  and survey.
\newblock \emph{Annual Reviews in Control}, 61:\penalty0 101052, 2026.

\bibitem[Ito and Wasa(2026)]{IW26a}
J.~Ito and Y.~Wasa.
\newblock Partial observation amplifies model mismatch in {MAP} estimation via
  information-curvature margins.
\newblock In \emph{Proceedings of the IEEE Conference on Decision and Control
  (CDC)}, 2026.
\newblock To appear. arXiv:2608.24550.

\bibitem[Mayne et~al.(2000)Mayne, Rawlings, Rao, and Scokaert]{MRRS00}
D.~Q. Mayne, J.~B. Rawlings, C.~V. Rao, and P.~O.~M. Scokaert.
\newblock Constrained model predictive control: {S}tability and optimality.
\newblock \emph{Automatica}, 36\penalty0 (6):\penalty0 789--814, 2000.

\bibitem[Pan et~al.(2026)Pan, Yi, Zhang, Manchester, Qu, and Shi]{PYZ+26}
C.~Pan, Z.~Yi, J.~Zhang, Z.~Manchester, G.~Qu, and G.~Shi.
\newblock Hybrid feedback sampling for sample-efficient model predictive
  control.
\newblock arXiv:2608.19443, 2026.

\bibitem[Rawlings et~al.(2017)Rawlings, Mayne, and Diehl]{RMD17}
J.~B. Rawlings, D.~Q. Mayne, and M.~M. Diehl.
\newblock \emph{Model Predictive Control: Theory, Computation, and Design}.
\newblock Nob Hill Publishing, Madison, WI, 2nd edition, 2017.

\bibitem[Schramm et~al.(2026)Schramm, Nguimatsia-Tiofack, Perrin-Gilbert,
  Toussaint, and Carpentier]{SNTPG+26}
F.~Schramm, F.~Nguimatsia-Tiofack, N.~Perrin-Gilbert, M.~Toussaint, and
  J.~Carpentier.
\newblock Variance-reduced model predictive path integral via quadratic model
  approximation.
\newblock In \emph{Proceedings of Robotics: Science and Systems (RSS)}, volume
  XXII, 2026.

\bibitem[Wang et~al.(2026)Wang, Sharifi, and Fazlyab]{WSF26}
J.~Wang, S.~Sharifi, and M.~Fazlyab.
\newblock Generalized model predictive path integral control as
  expectation--maximization.
\newblock \emph{IEEE Control Systems Letters}, 10:\penalty0 2119--2124, 2026.

\bibitem[Williams et~al.(2016)Williams, Drews, Goldfain, Rehg, and
  Theodorou]{WDG+16}
G.~Williams, P.~Drews, B.~Goldfain, J.~M. Rehg, and E.~A. Theodorou.
\newblock Aggressive driving with model predictive path integral control.
\newblock In \emph{Proceedings of the IEEE International Conference on Robotics
  and Automation (ICRA)}, pages 1433--1440, 2016.

\bibitem[Williams et~al.(2018)Williams, Drews, Goldfain, Rehg, and
  Theodorou]{WDG+18}
G.~Williams, P.~Drews, B.~Goldfain, J.~M. Rehg, and E.~A. Theodorou.
\newblock Information-theoretic model predictive control: {T}heory and
  applications to autonomous driving.
\newblock \emph{IEEE Transactions on Robotics}, 34\penalty0 (6):\penalty0
  1603--1622, 2018.

\bibitem[Yan and Devasia(2024)]{YD24}
L.~L. Yan and S.~Devasia.
\newblock Output-sampled model predictive path integral control ({o-MPPI}) for
  increased efficiency.
\newblock In \emph{Proceedings of the IEEE International Conference on Robotics
  and Automation (ICRA)}, pages 14279--14285, 2024.

\bibitem[Yi et~al.(2024)Yi, Pan, He, Qu, and Shi]{YPH+24}
Z.~Yi, C.~Pan, G.~He, G.~Qu, and G.~Shi.
\newblock {CoVO-MPC}: {T}heoretical analysis of sampling-based {MPC} and
  optimal covariance design.
\newblock In \emph{Proceedings of the Learning for Dynamics and Control
  Conference (L4DC)}, volume 242 of \emph{Proceedings of Machine Learning
  Research}, pages 1122--1135, 2024.

\bibitem[Yoon and Kim(2026{\natexlab{a}})]{YK26a}
H.-J. Yoon and H.~Kim.
\newblock Finite-sample closed-loop stability of model predictive path integral
  control for linear time-invariant systems.
\newblock arXiv:2607.04006, 2026{\natexlab{a}}.

\bibitem[Yoon and Kim(2026{\natexlab{b}})]{YK26b}
H.-J. Yoon and H.~Kim.
\newblock Stochastic stability of nonlinear {MPPI} via contraction theory and
  control {L}yapunov functions.
\newblock arXiv:2607.06945, 2026{\natexlab{b}}.

\bibitem[Yoon et~al.(2022)Yoon, Tao, Kim, Hovakimyan, and Voulgaris]{YTK+22}
H.-J. Yoon, C.~Tao, H.~Kim, N.~Hovakimyan, and P.~Voulgaris.
\newblock Sampling complexity of path integral methods for trajectory
  optimization.
\newblock In \emph{Proceedings of the American Control Conference (ACC)}, pages
  3482--3487, 2022.

\end{thebibliography}

\appendix
\section{Proofs}
\label{app:proofs}

\subsection{Cost and exact MPC}
\label{app:lifted}

Stacking the predicted state trajectory of Section~\ref{sec:analysis} gives
\begin{equation}
    \begin{bmatrix}x_1\\x_2\\\vdots\\x_T\end{bmatrix}=\Psi x+\Theta U,\qquad
    \Psi:=\begin{bmatrix}A\\A^2\\\vdots\\A^T\end{bmatrix},\qquad
    \Theta:=\begin{bmatrix}B&0&\cdots&0\\AB&B&\cdots&0\\\vdots&\vdots&\ddots&\vdots\\A^{T-1}B&A^{T-2}B&\cdots&B\end{bmatrix},
    \label{eq:app:pred}
\end{equation}
with $\Psi\in\R^{nT\times n}$ and $\Theta\in\R^{nT\times d}$.
With the stacked weight $\bar Q:=\mathrm{blkdiag}(Q,\dots,Q,P)\in\R^{nT\times nT}$, which has $T-1$ copies of $Q$, the cost~\eqref{eq:lq} is $J(x,U)=x^\top Qx+\|\mathcal R(x,U)\|^2$ with the residual
\begin{equation}
    \mathcal R(x,U):=\begin{bmatrix}\bar Q^{1/2}(\Psi x+\Theta U)\\(I_T\otimes R^{1/2})U\end{bmatrix}=HU+Lx,\quad
    H:=\begin{bmatrix}\bar Q^{1/2}\Theta\\I_T\otimes R^{1/2}\end{bmatrix},\quad
    L:=\begin{bmatrix}\bar Q^{1/2}\Psi\\0\end{bmatrix}.
    \label{eq:app:residual}
\end{equation}
Expanding the squared norm gives the quadratic function~\eqref{eq:quad} with
\begin{equation}
    M=H^\top H=\Theta^\top\bar Q\Theta+I_T\otimes R,\qquad N=H^\top L=\Theta^\top\bar Q\Psi,\qquad c(x)=x^\top(Q+L^\top L)x.
    \label{eq:app:MN}
\end{equation}
The matrix $M$ is positive definite because $\Theta^\top\bar Q\Theta\succeq0$ and $I_T\otimes R\succ0$.
The term $x^\top Qx$ does not depend on $U$, so the weights of MPPI and the minimizer are those of the cost $\|\mathcal R(x,U)\|^2$ of Section~\ref{sec:nonlinear}.

\phantomsection\label{app:riccati}%
By the backward Riccati iteration \citep[Section~1.3.3]{RMD17}, which stays at the terminal weight $P$, the minimizer of the cost~\eqref{eq:lq} is the control sequence of the regulator, $u_{\tau}=-\Klq(A-B\Klq)^{\tau}x$, with the minimum value $x^\top Px$ for every $T\ge1$.
The first input of the minimizer is therefore $-\Klq x$, that is, $E_0F=\Klq$.
Since the cost~\eqref{eq:quad} is quadratic with the curvature $M$ and attains the minimum value $x^\top Px$ at $U^\star(x)$, it also follows that
\begin{equation}
    J(x,U)=(U-U^\star(x))^\top M(U-U^\star(x))+x^\top Px.
    \label{eq:app:Jform}
\end{equation}

\subsection{Proof of Theorem~\ref{thm:closedloop}}
\label{app:thm1}

Substituting the mean $\mu_{t|t}=\Gamma\mu_{t|t-1}-\Lambda x_t$ into the plant $x_{t+1}=Ax_t+BE_0\mu_{t|t}$ and the warm start $\mu_{t+1|t}=S\mu_{t|t}$ gives~\eqref{eq:Acl}.

(a) The matrix $\Sigma_0M$ is similar to the positive definite matrix $\Sigma_0^{1/2}M\Sigma_0^{1/2}$, so $\Gamma=(I+\alpha\eta\Sigma_0M)^{-1}=W(I+\alpha\eta D)^{-1}W^{-1}$ for a fixed invertible $W$ and a diagonal $D\succ0$, and the entries of $\alpha\eta D$ are the eigenvalues $s_i$ of $\Sh$, which are proportional to $\eta$.
Hence $\Gamma\to I$ and $\Lambda=(I-\Gamma)F\to0$ as $\eta\to0$, and $\Gamma\to0$ and $\Lambda\to F$ as $\eta\to\infty$, which with $E_0F=\Klq$ (Appendix~\ref{app:riccati}) gives~\eqref{eq:limits}.
The limits are block triangular with the diagonal blocks $A$ and $S$, and $A-B\Klq$ and $0$, and $S$ is nilpotent, so their spectra are those of $A$ and of $A-B\Klq$ together with zeros.

(b) The spectral radius is continuous in the entries of a matrix, and $\calA(\eta\Sigma_0)$ converges to the limits~\eqref{eq:limits}.
Hence $\rho(\calA(\eta\Sigma_0))$ tends to $\rho(A)>1$ as $\eta\to0$ and to $\rho(A-B\Klq)<1$ as $\eta\to\infty$.
There thus exist $\eta_1>0$ with $\rho(\calA(\eta\Sigma_0))>1$ for all $\eta<\eta_1$ and $\eta_2$ with $\rho(\calA(\eta\Sigma_0))<1$ for all $\eta>\eta_2$, and $\eta_1\le\eta_2$ since no $\eta$ satisfies both inequalities. \jmlrQED

We finally justify the remark on the number of samples after Theorem~\ref{thm:closedloop}.
With $K$ samples, MPPI uses the estimate $\hat\mu_{t|t}$, which is the ratio of the sample means of $wU$ and $w$.
For each fixed state and memory, the strong law of large numbers applied to the numerator and the denominator shows that $\hat\mu_{t|t}\to\mu_{t|t}$ almost surely as $K\to\infty$.

\subsection{Proof of Proposition~\ref{prop:inference}}
\label{app:ess}

By~\eqref{eq:app:Jform}, the weight is $w=e^{-x^\top Px/\lambda}w_0$ with $w_0(U):=\exp(-(U-U^\star)^\top M(U-U^\star)/\lambda)$.
The first factor does not depend on the control sequence and cancels in the ratio~\eqref{eq:ess}, so $\ess=\E_p[w_0]^2/\E_p[w_0^2]$.
Under the prior $p$, the deviation $U-U^\star$ is distributed as $\mathcal N(e,\Sigma)$ with the gap $e:=\mu_{t|t-1}-U^\star$.
For a Gaussian vector $y\sim\mathcal N(e,\Sigma)$ and a matrix $G\succeq0$, completing the square gives
\begin{equation}
    \E\big[\exp(-y^\top Gy)\big]=\det(I+2\Sigma G)^{-1/2}\exp\big(-e^\top G(I+2\Sigma G)^{-1}e\big).
    \label{eq:app:gaussint}
\end{equation}
Applying~\eqref{eq:app:gaussint} with $y=U-U^\star$ and $G=jM/\lambda$ for $j=1,2$, and using $\det(I+j\alpha\Sigma M)=\det(I+j\Sh)$, gives $\E_p[w_0^j]=\det(I+j\Sh)^{-1/2}\exp\big(-\tfrac{j}{\lambda}e^\top M(I+j\alpha\Sigma M)^{-1}e\big)$.
In the ratio $\E_p[w_0]^2/\E_p[w_0^2]$, the determinant factors give $\prod_{i=1}^d\varphi(s_i)$ with $\varphi(s):=\sqrt{1+2s}/(1+s)=(1-\kappa^2)^{1/2}$ for $\kappa=s/(1+s)$, because $(1+2s)/(1+s)^2=1-(s/(1+s))^2$.
The exponential factors give the exponent $-\frac{2}{\lambda}e^\top[M(I+\alpha\Sigma M)^{-1}-M(I+2\alpha\Sigma M)^{-1}]e$.
With the identity $M(I+j\alpha\Sigma M)^{-1}=\alpha^{-1}\Sigma^{-1/2}\Sh(I+j\Sh)^{-1}\Sigma^{-1/2}$ and the whitened gap $\tilde e=\Sigma^{-1/2}e$, this exponent is $-\tilde e^\top[\Sh(I+\Sh)^{-1}-\Sh(I+2\Sh)^{-1}]\tilde e=-\tilde e^\top\Sh^2(I+\Sh)^{-1}(I+2\Sh)^{-1}\tilde e$, so that
\begin{equation}
    \ess=\prod_{i=1}^d(1-\kappa_i^2)^{1/2}\,\exp\!\big(-\tilde e^\top\Sh^2(I+\Sh)^{-1}(I+2\Sh)^{-1}\tilde e\big).
    \label{eq:app:essfull}
\end{equation}
We call the exponential in~\eqref{eq:app:essfull} the gap factor.
The matrices $\Sh^2$ and $(I+\Sh)^{-1}(I+2\Sh)^{-1}$ commute and are positive definite, so the exponent is nonpositive and vanishes if and only if $\tilde e=0$, which proves~\eqref{eq:essclosed} and the equality condition. \jmlrQED

We add two remarks used in Sections~\ref{sec:prelim} and~\ref{sec:analysis}.
Replacing $(\Sigma,\lambda)$ by $(c\Sigma,c\lambda)$ for $c>0$ leaves $\Sh$ unchanged and scales $\tilde e$ by $c^{-1/2}$, so the temperature at a fixed product $\alpha\Sigma$ changes only the gap factor.
Finally, the normalized ESS $1/(K\sum_k\bar w_k^2)=(K^{-1}\sum_kw(U^{(k)}))^2/(K^{-1}\sum_kw(U^{(k)})^2)$ converges almost surely to $\ess$ by the strong law of large numbers, since $0<w\le1$ by $J\ge0$.

\subsection{Mean of SS-MPPI and proof of Theorem~\ref{thm:collapsed}}
\label{app:collapsedmean}

Let $M_h$, $M_{hs}$, and $M_s$ be the blocks of $V^\top MV$, and let $N_h$ and $N_s$ be those of $V^\top N$.
The matrix $V^\top MV$ is positive definite, so its block $M_h$ and the reduced curvature $\Mb:=M_s-M_{hs}^\top M_h^{-1}M_{hs}$, which is the Schur complement of $M_h$, are positive definite.
Completing the square in $z_h$ in~\eqref{eq:quad} gives
\begin{equation}
    J(x,V_hz_h+V_sz_s)=(z_h-z_h^\star)^\top M_h(z_h-z_h^\star)+\bar J(x,z_s),\qquad
    \bar J(x,z_s)=z_s^\top\Mb z_s+2z_s^\top\Nb x+\bar c(x),
    \label{eq:app:hardsquare}
\end{equation}
with the minimizer $z_h^\star(x,z_s)=-M_h^{-1}(M_{hs}z_s+N_hx)$, the reduced cross term $\Nb:=N_s-M_{hs}^\top M_h^{-1}N_h$, and a term $\bar c(x)$ that does not depend on $z_s$.
The minimizer is affine in $z_s$, and $V_hz_h^\star(x,z_s)+V_sz_s=-L_hx+T_sz_s$ with the gain $L_h:=V_hM_h^{-1}N_h$ and the lifting $T_s:=V_s-V_hM_h^{-1}M_{hs}$.

Integrating $e^{-J/\lambda}$ over $z_h$ gives $(\pi\lambda)^{d_h/2}\det(M_h)^{-1/2}e^{-\bar J(x,z_s)/\lambda}$, whose first factor depends neither on $z_s$ nor on $x$.
Hence the density~\eqref{eq:collapsed} is integrable, its marginal in the sampled coordinates is proportional to $e^{-\bar J(x,z_s)/\lambda}\mathcal N(z_s;m_{t|t-1},\Sigma_s)$, and its conditional law of the computed coordinates given $z_s$ is $\mathcal N(z_h^\star(x,z_s),(\lambda/2)M_h^{-1})$.
With the reduced gain $F_s:=\Mb^{-1}\Nb$, the minimizer $-F_sx$ of $\bar J$ is the sampled part $V_s^\top U^\star(x)$ of the minimizer of $J$, since minimizing over $z_h$ and then over $z_s$ minimizes over both.
The reduced gain is thus $F_s=V_s^\top F$, which is the form used in Theorem~\ref{thm:collapsed}(a).
This marginal is therefore the inference of Section~\ref{sec:analysis} with $M$, $U^\star(x_t)$, $\Sigma$, and $\mu_{t|t-1}$ replaced by $\Mb$, $V_s^\top U^\star(x_t)$, $\Sigma_s$, and $m_{t|t-1}$, and~\eqref{eq:update} gives its mean
\begin{equation*}
    m_{t|t}=\Gamma_s m_{t|t-1}+(I-\Gamma_s)V_s^\top U^\star(x_t)=\Gamma_s m_{t|t-1}-\Lambda_sx_t,\qquad \Lambda_s:=(I-\Gamma_s)F_s.
\end{equation*}
The conditional mean $z_h^\star(x,z_s)$ of the computed coordinates is affine in $z_s$, so the tower property gives the mean control sequence
\begin{equation}
    U_{t|t}=V_h\E[z_h^\star(x_t,z_s)]+V_s m_{t|t}=-L_hx_t+T_s m_{t|t}.
    \label{eq:app:plan}
\end{equation}

\phantomsection\label{app:thm3}%
We now prove Theorem~\ref{thm:collapsed}.
(a) Substituting $m_{t|t}=\Gamma_s m_{t|t-1}-\Lambda_sx_t$ into~\eqref{eq:app:plan} gives $U_{t|t}=-(L_h+T_s\Lambda_s)x_t+T_s\Gamma_s m_{t|t-1}$, and stacking $x_{t+1}=Ax_t+BE_0U_{t|t}$ with $m_{t+1|t}=V_s^\top SU_{t|t}$ gives $\zeta_{t+1}=\calA_s(\Sigma_s)\zeta_t$ with
\begin{equation}
    \calA_s(\Sigma_s):=\begin{bmatrix}A-BE_0(L_h+T_s\Lambda_s) & BE_0T_s\Gamma_s\\ -V_s^\top S(L_h+T_s\Lambda_s) & V_s^\top ST_s\Gamma_s\end{bmatrix}.
    \label{eq:Asplit}
\end{equation}
As $\Sigma_s\to0$, $\Gamma_s\to I$ and $\Lambda_s\to0$, so $\calA_s(\Sigma_s)$ tends to $\calA_0$ of~\eqref{eq:A0}.
The spectral radius is continuous in the entries of a matrix, so if $\rho(\calA_0)<1$, there is $\varsigma>0$ with $\rho(\calA_s(\Sigma_s))<1$ for every $\Sigma_s\succ0$ with $\|\Sigma_s\|<\varsigma$, and the closed-loop system is then exponentially stable.

(b) The marginal above is the inference of Section~\ref{sec:analysis} with the replacements above, so the formula~\eqref{eq:app:essfull} gives
\begin{equation}
    \ess=\prod_{i=1}^{d_s}\varphi(\bar s_i)\,\exp\!\big(-\tilde e_s^\top\Sh_s^2(I+\Sh_s)^{-1}(I+2\Sh_s)^{-1}\tilde e_s\big),
    \label{eq:esscollapsed}
\end{equation}
with $\varphi$ of Appendix~\ref{app:ess}, the eigenvalues $\bar s_i$ of $\Sh_s$, and the whitened gap $\tilde e_s:=\Sigma_s^{-1/2}e_s$ of the gap $e_s:=m_{t|t-1}-V_s^\top U^\star(x_t)$.
As in Appendix~\ref{app:ess}, the exponent is nonpositive and vanishes if and only if $e_s=0$, which proves the bound and the equality condition of (b).
The matrices $\Sh_s^2$ and $(I+\Sh_s)^{-1}(I+2\Sh_s)^{-1}$ commute, and the norm of the latter is at most one, so the magnitude of the exponent is at most $\|\Sh_s\tilde e_s\|^2=\alpha^2e_s^\top\Mb\Sigma_s\Mb e_s\le\alpha^2\|\Mb\|^2\|\Sigma_s\|\,\|e_s\|^2$.
The gap factor therefore tends to one as $\Sigma_s\to0$ at a fixed $(x_t,m_{t|t-1})$, and so does the ESS, since every $\bar s_i\to0$, which proves the limit in (b). \jmlrQED

\subsection{Counterexample and proof of Proposition~\ref{prop:shift}}
\label{app:counter}

The directions of the largest curvature need not stabilize, as the following example shows.

\begin{example}
\label{ex:counter}
Let $T=1$, $\lambda=1$, $A=\mathrm{diag}(6/5,1/2)$, $B=R=I_2$, $Q=\mathrm{diag}(7/25,\,100-25/101)$, and $P=\mathrm{diag}(1,100)$, which is the stabilizing solution of the scalar equations $p=q+a^2p/(1+p)$ of the DARE, with $A-B\Klq=\mathrm{diag}(3/5,1/202)$.
Then $M=R+P=\mathrm{diag}(2,101)$, $N=PA=\mathrm{diag}(6/5,50)$, and $S=0$, so the spectrum of $\calA_0$ is that of $A-BE_0L_h$ with the zero of the memory.
With $\calH$ spanned by the second unit vector $e_2$, the direction of largest curvature, $L_h=V_hM_h^{-1}N_h$ has the single nonzero entry $50/101$, and $A-BE_0L_h=\mathrm{diag}(6/5,1/202)$ has the unstable eigenvalue $6/5$.
With $\calH$ spanned by the first unit vector $e_1$, the entry is $3/5$, and $A-BE_0L_h=\mathrm{diag}(3/5,1/2)$ is Schur.
\end{example}

\phantomsection\label{app:shift}%
We now prove Proposition~\ref{prop:shift}.
If $S\calH\subseteq\calH$, the columns of $SV_h$ lie in $\calH$, so $V_s^\top SV_h=0$.
Then the lower-left block of $\calA_0$ is $-V_s^\top SL_h=-V_s^\top SV_hM_h^{-1}N_h=0$ and $V_s^\top ST_s=V_s^\top SV_s$, so $\calA_0$ is block upper triangular and $\mathrm{spec}(\calA_0)=\mathrm{spec}(A-BE_0L_h)\cup\mathrm{spec}(V_s^\top SV_s)$.
In the basis $V$, the shift is block upper triangular with the diagonal blocks $V_h^\top SV_h$ and $V_s^\top SV_s$, so $(V_s^\top SV_s)^j$ is a block of $V^\top S^jV$, which vanishes for $j\ge T$.
Hence $V_s^\top SV_s$ is nilpotent and $\rho(\calA_0)=\rho(A-BE_0L_h)$ follows. \jmlrQED

We finally verify the condition for the subspace of Algorithm~\ref{alg:collapsed}.
The span of the first $k_b$ input blocks is shift-invariant because $S$ maps the input block $\tau$ to the block $\tau-1$ and the block $0$ to zero.

\subsection{Proof of Proposition~\ref{prop:local}}
\label{app:local}

We first verify the remark of Section~\ref{sec:nonlinear} on the LQ setting.
For a quadratic cost, the curvature block $M_h$ of the computed coordinates does not depend on $z_s$, so the factor $\det(M_h)^{-1/2}$ of Appendix~\ref{app:collapsedmean} is constant.
Weighting each sample by $e^{-J/\lambda}$ at the minimizer $z_h^\star$ therefore gives the mean of~\eqref{eq:collapsed}.

We now prove Proposition~\ref{prop:local}.
Write $\chi:=(x_t,\mu_{t|t-1})$ and let $H_h$ and $H_s$ be evaluated at $\chi$.
Near the origin, let $H_h^\dagger(\chi):=(H_h^\top H_h)^{-1}H_h^\top$, $g(\chi):=\mu_{t|t-1}-V_hH_h^\dagger(\chi)\mathcal R(x_t,\mu_{t|t-1})$, and $T_s(\chi):=V_s-V_hH_h^\dagger(\chi)H_s$.
The three maps are continuously differentiable near the origin, since $\mathcal R$ is of class $C^2$ and $H_h$ has full column rank at the origin.
With $\delta_s=\Sigma_s^{1/2}\nu$ and $\nu\sim\mathcal N(0,I_{d_s})$, the sample of Algorithm~\ref{alg:collapsed} is $G(\chi,\nu):=g(\chi)+T_s(\chi)\Sigma_s^{1/2}\nu$ by~\eqref{eq:gn}.
The controller with infinitely many samples applies the weighted mean
\begin{equation*}
    \bar U_{\Sigma_s}(\chi):=\frac{\E[e^{-J(x_t,G(\chi,\nu))/\lambda}G(\chi,\nu)]}{\E[e^{-J(x_t,G(\chi,\nu))/\lambda}]}.
\end{equation*}
The closed-loop system of Proposition~\ref{prop:local} is $\chi_{t+1}=\Phi_{\Sigma_s}(\chi_t)$ with the map
\begin{equation*}
    \Phi_{\Sigma_s}(\chi):=\big(f(x_t,E_0\bar U_{\Sigma_s}(\chi)),\,S\bar U_{\Sigma_s}(\chi)\big),
\end{equation*}
and $\Phi_0$ denotes the map with $\bar U_0:=g$.
At the origin $\mathcal R=0$, so the terms of the Jacobian of $g$ that contain derivatives of $H_h^\dagger$ vanish.
With the Jacobian $L$ of $\mathcal R$ with respect to $x$ at the origin, $M=H^\top H$ and $N=H^\top L$ give $H_h^\dagger L=M_h^{-1}N_h$ and $H_h^\dagger H_s=M_h^{-1}M_{hs}$.
Hence $\nabla_xg(0)=-V_hH_h^\dagger L=-L_h$, $\nabla_\mu g(0)V_s=V_s-V_hH_h^\dagger H_s=T_s$, and $\nabla_\mu g(0)V_h=V_h-V_hH_h^\dagger H_h=0$.
Let $\calA_{\mathrm{gn}}:=D\Phi_0(0)$ be the Jacobian in the coordinates $\chi$.
In the coordinates $(x_t,m_{t|t-1},h_{t|t-1})$ with $m_{t|t-1}:=V_s^\top\mu_{t|t-1}$ and $h_{t|t-1}:=V_h^\top\mu_{t|t-1}$, obtained from $\chi$ by an orthogonal change of coordinates, the Jacobian $\calA_{\mathrm{gn}}$ therefore becomes
\begin{equation*}
    \widetilde{\calA}_0:=\begin{bmatrix}A-BE_0L_h & BE_0T_s & 0\\ -V_s^\top SL_h & V_s^\top ST_s & 0\\ -V_h^\top SL_h & V_h^\top ST_s & 0\end{bmatrix},
\end{equation*}
which is block triangular with the diagonal blocks $\calA_0$ and $0$, so $\rho(\calA_{\mathrm{gn}})=\rho(\widetilde{\calA}_0)=\rho(\calA_0)$.
If $\rho(\calA_0)<1$, let $P_c\succ0$ solve $\calA_{\mathrm{gn}}^\top P_c\calA_{\mathrm{gn}}-P_c=-I$, and write $\|\chi\|_{P_c}:=(\chi^\top P_c\chi)^{1/2}$ and $\|X\|_{P_c}:=\|P_c^{1/2}XP_c^{-1/2}\|$.
The precise claim is that, if $\rho(\calA_0)<1$, there are $r_c>0$, $\varsigma>0$, and $a_1<1$ with the following property for every $\Sigma_s\succ0$ with $\|\Sigma_s\|<\varsigma$.
The map $\Phi_{\Sigma_s}$ has a unique fixed point $\chi^\star$ with $\|\chi^\star\|_{P_c}\le r_c$, and every trajectory with $\|\chi_0\|_{P_c}\le r_c$ satisfies $\|\chi_t-\chi^\star\|_{P_c}\le a_1^t\|\chi_0-\chi^\star\|_{P_c}$.
Moreover, $\chi^\star\to0$ as $\Sigma_s\to0$.
In particular, $\|\chi_t\|_{P_c}\le\|\chi^\star\|_{P_c}+a_1^t\|\chi_0-\chi^\star\|_{P_c}$, so every trajectory from the ball converges exponentially to the ball of radius $\|\chi^\star\|_{P_c}$ around the origin, which shrinks to the origin as $\Sigma_s\to0$.

\begin{proof}
On a compact neighborhood of the origin and for $\|\Sigma_s\|\le1$, the sample $G$ and its Jacobian in $\chi$ grow at most linearly in $\|\nu\|$, since $G$ is affine in $\nu$ with continuous coefficients.
Since $0<e^{-J/\lambda}\le1$ and $J$ and $\nabla J$ grow at most polynomially in $U$, the integrands of the numerator and the denominator of $\bar U_{\Sigma_s}$ and their derivatives in $\chi$ are bounded by a common polynomial in $\|\nu\|$.
The polynomial is integrable under the Gaussian of $\nu$.
By dominated convergence, the numerator, the denominator, and their derivatives in $\chi$ are continuous in $(\chi,\Sigma_s^{1/2})$ on the compact set.
Hence the three converge uniformly in $\chi$ as $\Sigma_s\to0$ to their values at $\Sigma_s=0$, where $G=g(\chi)$.
The denominator at $\Sigma_s=0$ is $e^{-J(x_t,g(\chi))/\lambda}$, which has a positive minimum on the compact set, so $\bar U_{\Sigma_s}\to g$ and $D\Phi_{\Sigma_s}\to D\Phi_0$ uniformly on the compact set.
Let $a_0:=\|\calA_{\mathrm{gn}}\|_{P_c}<1$, choose $\ell\in(0,1-a_0)$, and let $a_1:=a_0+\ell$.
By the uniform convergence, the continuity of $D\Phi_0$, and $\Phi_0(0)=0$, there are $r_c>0$ and $\varsigma>0$ such that $\|D\Phi_{\Sigma_s}(\chi)-\calA_{\mathrm{gn}}\|_{P_c}\le\ell$ on the ball $\|\chi\|_{P_c}\le r_c$ and $\|\Phi_{\Sigma_s}(0)\|_{P_c}\le(1-a_1)r_c$ for $\|\Sigma_s\|<\varsigma$.
Then $\Phi_{\Sigma_s}$ is a contraction of the ball into itself with the constant $a_1$, and the Banach fixed-point theorem gives the claim, with $\|\chi^\star\|_{P_c}\le\|\Phi_{\Sigma_s}(0)\|_{P_c}/(1-a_1)$.
\end{proof}

\section{Inverted pendulum}
\label{app:linear}

\paragraph{Model.}
The inverted pendulum on a cart of Section~\ref{sec:numerical} is the benchmark of \citet{BSA83} without friction, with the equations of motion of \citet{Florian07}.
Let $p$ be the cart position, $\theta$ the pole angle, which is zero upright, and $u$ the horizontal force on the cart.
The equations of motion are then
\begin{equation}
    (m_c+m_p)\ddot p+m_pl\,(\ddot\theta\cos\theta-\dot\theta^2\sin\theta)=u,\qquad
    \tfrac43l\,\ddot\theta+\ddot p\cos\theta=g\sin\theta,
    \label{eq:app:cartpole}
\end{equation}
with the parameters of Table~\ref{tab:app:pendulum}.
We take the state $x:=(p,\dot p,\theta,\dot\theta)\in\R^4$, so that~\eqref{eq:app:cartpole} defines the state equation $\dot x=f_c(x,u)$.
Linearizing the state equation at the upright equilibrium $(x,u)=(0,0)$ gives
\begin{equation}
    \dot x=A_cx+b_cu,\qquad
    A_c=\begin{bmatrix}0&1&0&0\\0&0&a_{23}&0\\0&0&0&1\\0&0&a_{43}&0\end{bmatrix},\qquad
    b_c=\begin{bmatrix}0\\b_2\\0\\b_4\end{bmatrix},
    \label{eq:app:pendlin}
\end{equation}
with $a_{23}=-3m_pg/(4m_c+m_p)$, $a_{43}=3(m_c+m_p)g/(l(4m_c+m_p))$, $b_2=4/(4m_c+m_p)$, and $b_4=-3/(l(4m_c+m_p))$.
We next discretize the state equation.
The model $f$ of~\eqref{eq:ocp} is the fourth-order Runge--Kutta step of $f_c$ with the period $\Delta t$.
The matrices $A$ and $B$ are thus the zero-order-hold discretization of~\eqref{eq:app:pendlin} truncated at the fourth order, and $\rho(A)=1.0827\approx e^{\sqrt{a_{43}}\Delta t}$.
The cost is~\eqref{eq:lq} along the predicted state trajectory of $f$ with the weights of Table~\ref{tab:app:pendulum}.
The terminal weight $P$ is set to the solution of the DARE for $(A,B,Q,R)$, for which $\rho(A-B\Klq)=0.9735$.

\paragraph{Thresholds and Monte Carlo estimates.}
For CoVO-MPC \citep{YPH+24}, we take the covariance $\Sigma=\sigma^2\det(M)^{1/(2d)}M^{-1/2}$, which has the determinant of $\sigma^2I$.
We compute the thresholds of Theorem~\ref{thm:closedloop} on a grid of $401$ points in $\sigma^2$ and refine each threshold by bisection.
The thresholds are $\sigma^2=0.0327$ for MPPI and $0.327$ for CoVO-MPC.
Both spectral radii cross one exactly once on the grid, so the stable sets are intervals there.
SS-MPPI has no threshold.
In the same way, we compute the variance at which the ESS of MPPI at zero gap falls below one half, which is $\sigma^2=2.0$.
For the Monte Carlo estimates of Figure~\ref{fig:cartpole}(b), we use $K=4096$ samples and take the mean over ten repetitions.
At the seven variances checked, the largest relative error from~\eqref{eq:essclosed} is $0.95$ percent.

\paragraph{Severity of the conflict.}
We finally examine how severe the conflict of Section~\ref{sec:analysis} is on other plants.
We take plants with one input, $Q=I$, $R=1$, $\lambda=1$, $T=10$, and $\Sigma=\sigma^2I$ (Table~\ref{tab:app:severity}).
The thresholds and the variances are computed as above, with the scan from $10^{-12}$ to $10^{2}$.
For the scalar plant $x_{t+1}=ax_t+u_t$, the bound~\eqref{eq:essclosed} at the threshold of Theorem~\ref{thm:closedloop} falls from $0.996$ to $0.745$ as the pole $a$ moves from $1.1$ to $3$.
The bound at the threshold thus stays above $0.7$ for these poles, and the pendulum, with the pole $1.08$, has the bound $0.996$ at the threshold.
For the two-pole plants, in contrast, the conflict is severe, with the bound at the threshold between $0.0004$ and $0.072$.

\begin{table}[t]
\centering
\caption{Parameters of the inverted pendulum. The mechanical parameters and the period follow \citet{BSA83}, except $g=9.81$ instead of $9.8$ m/s$^2$ and the Runge--Kutta step instead of the Euler step.}
\label{tab:app:pendulum}
\begin{tabular}{llc}
\hline
symbol & description & value \\
\hline
$m_c$ & mass of the cart & $1$ kg \\
$m_p$ & mass of the pole & $0.1$ kg \\
$l$ & half-length of the pole & $0.5$ m \\
$g$ & gravitational acceleration & $9.81$ m/s$^2$ \\
$\Delta t$ & sampling period & $0.02$ s \\
$Q$, $R$ & weights of the cost~\eqref{eq:lq} & $\mathrm{diag}(1,0.1,10,0.1)$, $0.01$ \\
$T$, $\lambda$ & horizon and temperature & $20$, $1$ \\
$x_0$ & initial state in Figure~\ref{fig:cartpole}(c) & $(0,0,0.05,0)$ \\
\hline
\end{tabular}
\end{table}

\begin{table}[t]
\centering
\caption{Severity of the conflict for plants with one input, $Q=I$, $R=1$, $\lambda=1$, $T=10$, and $\Sigma=\sigma^2I$: the poles of $A$, the threshold of Theorem~\ref{thm:closedloop}, the bound~\eqref{eq:essclosed} at the threshold, and the variance at which the bound falls to one half. The two-pole plants have $B=(1,1)^\top$.}
\label{tab:app:severity}
\begin{tabular}{cccc}
\hline
poles of $A$ & threshold $\sigma^2$ & bound at the threshold & $\sigma^2$ at the bound $1/2$ \\
\hline
$1.1$ & $3.8\times10^{-4}$ & $0.996$ & $2.4\times10^{-2}$ \\
$1.5$ & $2.7\times10^{-5}$ & $0.943$ & $3.5\times10^{-4}$ \\
$2$ & $3.1\times10^{-7}$ & $0.866$ & $2.0\times10^{-6}$ \\
$3$ & $2.5\times10^{-10}$ & $0.745$ & $8.0\times10^{-10}$ \\
\hline
$1.2$, $1.5$ & $1.5\times10^{-3}$ & $0.072$ & $2.5\times10^{-5}$ \\
$1.5$, $2$ & $3.1\times10^{-4}$ & $0.0085$ & $7.9\times10^{-8}$ \\
$1.5$, $3$ & $1.4\times10^{-4}$ & $0.0004$ & $8.0\times10^{-11}$ \\
\hline
\end{tabular}
\end{table}

\section{Quadruped robot}
\label{app:go2}

\paragraph{Model.}
The quadruped of Section~\ref{sec:numerical} is the Unitree Go2 in the Genesis simulator \citep{Genesis24}, with the hip, thigh, and calf joints in each leg.
The state $x\in\R^{48}$ consists of the base position, the base orientation, the base linear and angular velocities in the body frame, the joint angles $q\in\R^{12}$, the joint velocities $\dot q\in\R^{12}$, and the previous input.
Here, the base orientation is a rotation vector relative to the standing orientation.
The input $u\in\R^{12}$ shifts the targets of the joint PD loops.
During the time step $t$, the PD loops apply the torques $k_p(q^{\mathrm{def}}+0.25\,u_{t-1}-q)-k_d\dot q$ with the standing posture $q^{\mathrm{def}}$ and the gains of Table~\ref{tab:app:go2}.
The torques thus depend on the previous input $u_{t-1}$, which is why the state contains the previous input.
The model $f$ in~\eqref{eq:ocp} is then a first-order recursion despite the one-step latency.
The model advances the state by the period $\Delta t=0.02$ s in two substeps of the simulator and is time-invariant.
The contact forces come from the constraint solver of the simulator.

\paragraph{Cost.}
We next specify the cost.
The cost is the sum of the squared stage residuals $\|r(x_\tau)\|^2$ over $\tau=1,\dots,T$ along the predicted state trajectory of $f$ and a terminal cost.
The terminal cost is $(x_T-x^{\mathrm{ref}})^\top P_T(x_T-x^{\mathrm{ref}})$, where the weight $P_T$ and the $45$ coordinates that enter are specified below.
The reference $x^{\mathrm{ref}}$ is the standing state $x^{\mathrm{st}}$ defined below with the base linear velocity replaced by $(v^{\mathrm{cmd}},0)$ and the angular velocity by $(0,0,\omega^{\mathrm{cmd}})$.
With the references and the weights of Table~\ref{tab:app:go2}, the stage residual $r(x)\in\R^{30}$ stacks
\begin{equation*}
    \sqrt{w_v}\,(v_{xy}-v^{\mathrm{cmd}}),\enspace \sqrt{w_\omega}\,(\omega_z-\omega^{\mathrm{cmd}}),\enspace \sqrt{w_h}\,(h-h^{\mathrm{ref}}),\enspace \sqrt{w_g}\,g_{xy},\enspace \sqrt{w_q}\,(q-q^{\mathrm{def}}),\enspace \sqrt{w_u}\,u^-.
\end{equation*}
Here, $v_{xy}$ is the forward and lateral base velocity in the body frame, $\omega_z$ the yaw rate, and $h$ the base height.
The vector $g_{xy}$ is the horizontal part of the gravity direction in the body frame, which vanishes when the base is level.
The vector $u^-$ is the previous input in the state.
The residual $\mathcal R$ of Section~\ref{sec:nonlinear} stacks $r(x_1),\dots,r(x_T)$ and the terminal rows $P_T^{1/2}(x_T-x^{\mathrm{ref}})$.

\paragraph{Linearization and design.}
We next design the stabilizing subspace and the terminal weight from a linearization of the model $f$.
We first simulate the robot for $6$ s from the standing posture with $u=0$ and call the resulting state the standing state $x^{\mathrm{st}}$.
We then differentiate $f$ and the stage residual $r$ at $(x^{\mathrm{st}},0)$ by central differences with the step $10^{-4}$.
Smaller steps suffer from the single-precision rounding of the simulator, and larger steps make or break foot contacts.
On flat ground, the base position in the plane and the yaw affect neither the other coordinates nor the residual.
We thus drop the three coordinates and obtain the matrices $A\in\R^{45\times45}$ and $B\in\R^{45\times12}$ of Algorithm~\ref{alg:collapsed}.
Let $C\in\R^{30\times45}$ denote the Jacobian of the stage residual $r$ with respect to the state.
The terminal weight is set to $P_T=P-C^\top C$, where $P$ solves the DARE for $(A,B,C^\top C,R)$ with $R=10^{-9}w_uI$.
Here, the weight $R$ is small because the input is already weighted in $C^\top C$ through the previous input.
The term $C^\top C$ is subtracted because the stage residual at $x_T$ is already counted.
The linearized model has $\rho(A)=1.0013$, so the standing robot is close to marginally stable, and $\rho(A-B\Klq)=0.996$.
We finally apply the rule of Algorithm~\ref{alg:collapsed}.
The first input block alone gives $\rho(A-BE_0L_h)=1.027$, which is less stable than the plant, whereas the first two input blocks give $0.9959$.
The rule thus selects $k_b=2$, that is, $d_h=24$ computed coordinates.

\paragraph{Experimental settings.}
All controllers use the parameters of Table~\ref{tab:app:go2} unless stated otherwise.
Every run starts from the same settled standing pose with a zero warm start, and the seed sets only the sampling noise of the controller.
A run lasts $600$ time steps ($12$ s) and ends early at a fall.
We count a fall when $|\mathrm{roll}|$ or $|\mathrm{pitch}|$ exceeds $60^\circ$ or the base height drops below $0.15$ m.
A run without a fall is classified as walking if the mean forward velocity exceeds $0.1$ m/s and as standing otherwise.
One run of CoVO-MPC at $\sigma^2=0.1$ falls after $10$ s, and no other run falls.

We next describe the computation of the controllers.
SS-MPPI computes the Jacobian $H$ of~\eqref{eq:gn} by central differences with the step $10^{-2}$ from one batch of $241$ rollouts, the nominal sequence and two per coordinate.
CoVO-MPC computes the curvature in the same way as the Gauss--Newton approximation $M_t=H^\top H$ without regularization.
SS-MPPI with $k_b=T$ applies the step~\eqref{eq:gn} to all $d=120$ coordinates with the same Jacobian and terminal weight, without sampling.
One time step of SS-MPPI or CoVO-MPC thus takes about twice the time of MPPI ($29$ ms against $14$ ms on an RTX 5090).
Every rollout batch is padded to $2048$ copies of the robot, so the times hold for every $K$ from $16$ to $1024$.
The simulator advances only after each control update, so the runs impose no real-time deadline on the controller.

We finally define the measures of Section~\ref{sec:numerical}.
The distance is the displacement of the base along the initial heading at the end of the run.
The pitch of Table~\ref{tab:go2} is the maximum of $|\mathrm{pitch}|$ over the run.
The ESS is the normalized ESS $1/(K\sum_k\bar w_k^2)$ along the run with $K=1024$, averaged over the time steps of the run and the five seeds.

\begin{table}[t]
\centering
\caption{Parameters of the Go2. The PD gains, the factor $0.25$ of the input, and the standing posture follow the locomotion example of Genesis \citep{Genesis24}.}
\label{tab:app:go2}
\begin{tabular}{llc}
\hline
symbol & description & value \\
\hline
$\Delta t$ & control period & $0.02$ s \\
$k_p$, $k_d$ & PD gains & $20$, $0.5$ \\
$q^{\mathrm{def}}$ & standing posture & hip $0$, thigh $0.8$ (rear $1.0$), calf $-1.5$ rad \\
$v^{\mathrm{cmd}}$, $\omega^{\mathrm{cmd}}$, $h^{\mathrm{ref}}$ & commands, reference height & $(0.5,0)$ m/s, $0$, $0.3$ m \\
$w_v$, $w_\omega$, $w_h$ & weights of the residual & $1$, $0.2$, $50$ \\
$w_g$, $w_q$, $w_u$ & weights of the residual & $5$, $0.1$, $0.005$ \\
$T$, $\lambda$, $K$ & horizon, temperature, samples & $10$, $0.03$, $1024$ \\
\hline
\end{tabular}
\end{table}

\end{document}